\documentclass[superscriptaddress]{revtex4}
\usepackage{amsfonts}
\usepackage{amsmath}
\usepackage{amssymb}

\input{tcilatex}
\begin{document}

\author{Mario Castagnino}
\address{CONICET, IAFE, Instituto de F\'{\i}sica de Rosario, Facultad de
Ciencias Exactas y Naturales, Universidad de Buenos Aires. Casilla de
Correos 67, Sucursal 28, 1428, Buenos Aires, Argentina.}
\author{Sebastian Fortin}
\affiliation{CONICET, IAFE, Facultad de Ciencias Exactas y Naturales, Universidad de
Buenos Aires. Casilla de Correos 67, Sucursal 28, 1428, Buenos Aires,
Argentina. e-mail: sfortin@gmx.net}
\author{Roberto Laura}
\affiliation{Departamento de F\'{\i}sica y Qu\'{\i}mica, Facultad de Ciencias Exactas,
Ingenier\'{\i}a y Agrimensura, Universidad Nacional de Rosario. Pellegrini
250, 2000, Rosario, Argentina.}
\author{Olimpia Lombardi}
\affiliation{CONICET, Universidad de Buenos Aires. Cris\'{o}logo Larralde 3440, 1430,
Buenos Aires, Argentina}
\title{A general theoretical framework for decoherence in open and closed
systems}

\begin{abstract}
A general theoretical framework for decoherence is proposed, which
encompasses formalisms originally devised to deal just with open or with
closed systems. The conditions for decoherence are clearly stated and the
relaxation and decoherence times are compared. Finally, the spin-bath model
is developed in detail from the new perspective.
\end{abstract}

\maketitle

\section{Introduction}

The specific feature of quantum mechanics is the superposition principle,
which leads to the phenomenon of interference. Therefore, any attempt to
account for the emergence of classicality from quantum behavior must explain
how interference vanishes. The process that cancels interference and selects
the candidates for classical states is usually called \textit{decoherence}.

>From a diachronic perspective, the decoherence program finds its roots
(though, of course, not under this name) in the attempts to explain how a
coherent pure state becomes a final decohered mixture with no interference
terms. Three general periods can be identified in the development of this
program:

\begin{itemize}
\item \textbf{Closed systems period} (van Kampen \cite{van Kampen}, van Hove 
\cite{van Hove-1}, \cite{van Hove-2}, Daneri \textit{et al.} \cite{Daneri}).
In order to understand how classical macroscopic features arise from quantum
microscopic behavior, ``gross'' observables are defined. The states,
indistinguishable for a macroscopic observer, are described by the same
coarse-grained state $\rho _{G}(t)$. When the evolution of $\rho _{G}(t)$
(or of the expectation value of the gross observables) is studied, it can be
proved that $\rho _{G}(t)$ approaches a final stable state through a process
with characteristic time $t_{C}$; therefore, $\rho _{G}(t)$ decoheres in its
own eigenbasis after a decoherence time $t_{D}=t_{C}$. The main problem of
this period was the fact that $t_{C}$ turned out to be too long to account
for experimental data (see \cite{Omnes-2005}).

\item \textbf{Open systems period}. An open system $S$ is considered in
interaction with its environment $E$, and the evolution of the reduced state 
$\rho _{S}(t)=Tr_{E}\rho _{SE}(t)$ is studied. The so-called \textit{%
environment induced decoherence} (EID) approach (Zeh \cite{Zeh-1970}, \cite%
{Zeh-1973}, Zurek \cite{Zurek-1982}, \cite{Zurek-1993}, \cite{Paz-Zurek}, 
\cite{Zurek-2003}) proves that, since the interference terms of $\rho
_{S}(t) $ rapidly vanish, $\rho _{S}(t)$ decoheres in an adequate pointer
basis after an extremely short decoherence time $t_{D}=t_{DS}$. This result
overcomes the main problem of the first period.

However, the foundations of the EID program are still threatened by certain
conceptual problems derived from its open-system perspective:

\begin{itemize}
\item \textit{The closed-system problem}: If only open systems may decohere,
the issue of the emergence of classicality in closed systems, in particular,
in the Universe as a whole, cannot even be posed (see Zurek in \cite%
{Zurek-1994}; for criticisms, see \cite{Pessoa}).

\item \textit{The defining systems problem}: Since the environment may be
external or internal, there is no general criterion to decide where to place
the ``cut'' between system and environment (see Zurek's formulation of this
problem in \cite{Zurek-1998}, and a discussion in \cite{CO-Studies}).
\end{itemize}

\item \textbf{Closed and open systems period.} Although at present EID is
still considered the ``orthodox'' view about the matter (\cite{Leggett}, 
\cite{Bub}), in the last times other approaches have been proposed to face
the conceptual difficulties of EID, in particular, the closed-system problem
(Diosi \cite{Diosi-1}, \cite{Diosi-2}, Milburn \cite{Milburn}, Penrose \cite%
{Penrose}, Casati and Chirikov \cite{Casati-Chirikov-1}, \cite%
{Casati-Chirikov-2}, Adler \cite{Adler}). Some of these methods are clearly
``non-dissipative'' (Bonifacio \textit{et al. }\cite{Bonifacio}, Ford and
O'Connell \cite{Ford}, Frasca \cite{Frasca}, Sicardi Schifino \textit{et al.}
\cite{Sicardi}). Among them, we have developed the \textit{self-induced
decoherence }(SID) approach, according to which a closed quantum system with
continuous spectrum may decohere by destructive interference, and may reach
a decohered state where the classical limit can be obtained (\cite%
{CO-Studies}, \cite{CL-PRA-2000}, \cite{CL-IJTP-2000}, \cite{CO-IJTP}, \cite%
{Cast-2004}, \cite{COrd}, \cite{Cast-2005}, \cite{CO-PR(DT)}, \cite{CO-PS}, 
\cite{Cast-2006}, \cite{CG-CL}, \cite{CO-CSF}).
\end{itemize}

Although the conceptual problems of EID may be not obstacles to local
applications of the theory, they turn out to be serious challenges in
cosmology, where the purpose is to explain the classical behavior of the
Universe, which, by definition, has no external environment. The usual
strategy in cosmology consists in splitting the Universe into some degrees
of freedom which represent the ``system'' of interest, and the remaining
degrees of freedom that are supposed to be non accessible and, therefore,
play the role of an internal environment. For instance, in quantum field
theory, when it is known that the background field follows a simple
classical behavior, the scalar field is decomposed according to $\phi =\phi
_{c}+\phi _{q}$, where the background field $\phi _{c}$ plays the role of
the system and the fluctuation field $\phi _{q}$ plays the role of the
environment (see \cite{Calzetta}). This means that the observables which
will behave classically must be assumed in advance: there is no general
criterion to discriminate between system and environment. This explains the
search for an account of decoherence in closed systems applicable to
cosmology. For instance, in \cite{CO-IJTP} we attempted to explain the
emergence of classicality in a Robertson-Walker Universe from the
perspective of the SID approach.

In spite of the fact that, at present, formalisms for closed and open
systems coexist, in the literature both kinds of approaches are often
presented as alternative scenarios for decoherence, or even as theories
dealing with different physical phenomena (\cite{Max}). In the next sections
we will challenge this common view by showing that both approaches can be
understood in the context of a general theoretical framework.

\section{A general framework for decoherence}

As emphasized by Omn\`{e}s (\cite{Omnes-2001}, \cite{Omnes-2002}),
decoherence is a particular case of the phenomenon of irreversibility, which
leads to the following problem. Since the quantum state $\rho (t)$ evolves
unitarily, it cannot follow an irreversible evolution. Therefore, if the
non-unitary evolution is to be accounted for, a further element has to be
added, precisely, the splitting of the maximal information about the system
into a relevant part and an irrelevant part: whereas the irrelevant part is
discarded, the relevant part may evolve non-unitarily. This idea can be
rephrased in operators language. The maximal information about the system is
given by the set $\mathcal{O}$ of all its possible observables. Therefore,
by selecting a subset $\mathcal{O}^{R}\subset \mathcal{O}$, we restrict that
maximal information to a relevant part.

Since decoherence is an irreversible process, the splitting of the whole set 
$\mathcal{O}$ of observables is also required. On this basis, the phenomenon
of decoherence can be explained in three general steps:

\begin{itemize}
\item \textbf{Step 1}: The set $\mathcal{O}^{R}\subset \mathcal{O}$ of
relevant observables is defined.

\item \textbf{Step 2}: The expectation value $\langle O^{R}\rangle _{\rho
(t)}$ is computed, for any $O^{R}\in \mathcal{O}^{R}$.

\item \textbf{Step 3}: It is proved that $\langle O^{R}\rangle _{\rho (t)}$
rapidly approaches a value $\langle O^{R}\rangle _{\rho _{*}}$ (or that $%
\rho (t)$ weakly approaches a final state $\rho _{*}$)
\end{itemize}

Since always a coarse-grained state $\rho _{G}(t)$ can be defined, such that 
$\langle O^{R}\rangle _{\rho (t)}=\langle O^{R}\rangle _{\rho _{G}(t)}$, the
non-unitary evolution of $\rho _{G}(t)$ (governed by a master equation) can
be obtained: $\rho _{G}(t)$ will rapidly converge to a final state $\rho
_{G*}$, which is obviously diagonal in its own eigenbasis: 
\begin{equation}
\langle O^{R}\rangle _{\rho (t)}=\langle O^{R}\rangle _{\rho
_{G}(t)}\longrightarrow \langle O^{R}\rangle _{\rho _{*}}=\langle
O^{R}\rangle _{\rho _{G*}}  \label{2-1}
\end{equation}
This means that, although the off-diagonal terms of $\rho (t)$ never vanish
through the unitary evolution, decoherence obtains because it is a \textit{%
coarse-grained process}: the system decoheres \textit{from the observational
point of view} given by any relevant observable $O^{R}\in \mathcal{O}^{R}$.

\section{Open and closed systems}

The need of selecting a set $\mathcal{O}^{R}$ of relevant observables, in
terms of which the time-evolution of the system is described, is explicitly
or implicitly admitted by the different approaches to the emergence of
classicality: gross observables in van Kampen \cite{van Kampen}, macroscopic
observables of the apparatus in Daneri \textit{et al.} \cite{Daneri},
observables of the open system in EID \cite{Zeh-1973}, \cite{Zurek-2003},
collective observables in Omn\`{e}s \cite{Omnes-1994}, \cite{Omnes-1999},
van Hove observables in SID \cite{CO-PR(DT)}, \cite{Cast-2006}, etc. It is
quite clear that a closed system can be ``partitioned'' into many different
ways and, thus, there is not a single set of relevant observables
essentially privileged (see \cite{Sujeeva-1}, \cite{Sujeeva-2}).
Nevertheless, when the emergence of classicality has to be accounted for,
certain sets $\mathcal{O}^{R}$ prove to be physically relevant, in the sense
that the expectation values $\langle O^{R}\rangle _{\rho (t)}$ follow the
non-unitary evolution required in Step 3, under particular definite
conditions that have to be defined in each particular case.

Here we will analyze two ways of selecting the relevant observables, those
proposed by the EID and the SID approaches, and we will study the conditions
that lead to decoherence in each case. In particular, we will show that both
approaches are not conflicting views, but they can be subsumed under the
general framework sketched in the previous section.

\subsection{Open systems}

EID is usually conceived as an open-systems approach because it partitions a
closed system $U$ into a proper system $S$ and its environment $E$. However,
as we will see, this approach can be rephrased from the viewpoint of the
system $U$ in the general framework introduced in the previous section.

Let us consider the Hilbert space $\mathcal{H}$ of the system $U$, $\mathcal{%
H}=\mathcal{H}_{S}\otimes $ $\mathcal{H}_{E}$, where $\mathcal{H}_{S}$ is
the Hilbert space of $S$ and $\mathcal{H}_{E}$ the Hilbert space of $E$. The
corresponding von Neumann-Liouville space of $U$ is $\mathcal{L}=\mathcal{%
H\otimes H=L}_{S}\otimes $ $\mathcal{L}_{E}$, where $\mathcal{L}_{S}=%
\mathcal{H}_{S}\otimes $ $\mathcal{H}_{S}$ and $\mathcal{L}_{E}=\mathcal{H}%
_{E}\otimes $ $\mathcal{H}_{E}$.

\textbf{Step 1:} In the case of EID, the relevant observables are those
corresponding to the open system $S$: 
\begin{equation}
O^{R}=O_{S}\otimes I_{E}\in \mathcal{O}^{R}\subset \mathcal{L}  \label{3-1}
\end{equation}
where $O_{S}\in \mathcal{L}_{S}$ and $I_{E}$ is the identity operator in $%
\mathcal{L}_{E}$.

\textbf{Step 2: }The expectation value of any $O^{R}\in \mathcal{O}^{R}$ in
the state $\rho (t)$ of $U$ can be computed as 
\begin{equation}
\langle O^{R}\rangle _{\rho (t)}=Tr\,\left( \rho (t)(O_{S}\otimes
I_{E})\right) =Tr\left( \rho _{S}(t)\,O_{S}\right) =\langle O_{S}\rangle
_{\rho _{S}(t)}  \label{3-2}
\end{equation}
where $\rho _{S}(t)=Tr_{E}\,\rho (t)$ is the reduced density operator of $S$%
, obtained by tracing over the environmental degrees of freedom.

\textbf{Step 3: }The EID approach studies the time evolution of $\rho _{S}(t)
$ governed by an effective master equation; it proves that, under certain
definite conditions, $\rho _{S}(t)$ converges to a stable state $\rho _{S*}$%
: 
\begin{equation}
\rho _{S}(t)\longrightarrow \rho _{S*},\qquad then\quad \langle O^{R}\rangle
_{\rho (t)}=\langle O_{S}\rangle _{\rho _{S}(t)}\longrightarrow \langle
O_{S}\rangle _{\rho _{S*}}=\langle O^{R}\rangle _{\rho _{*}}  \label{3-3}
\end{equation}
where $\rho _{S*}$ is obviously diagonal in its own eigenbasis. This process
of convergence of $\rho _{S}(t)$ to a final stable case has a characteristic
time $t_{RS}$, called \textit{relaxation time of }$S$, that is, the time
that the system needs to reach a state very close to the decohered
equilibrium state.

In the EID approach, another relevant time can be defined. Let us consider
the following states: 
\begin{equation}
|\Psi (0)\rangle =|\psi _{S}(0)\rangle \,|\psi _{E}(0)\rangle ,\qquad \text{ 
}|\psi _{S}(0)\rangle =\sum_{i}a_{i}|\psi _{iS}(0)\rangle  \label{3-31}
\end{equation}
where $|\psi _{S}(0)\rangle $ is the initial state of the proper system, $%
|\psi _{E}(0)\rangle $\ is the initial state of the environment, and $%
\{|\psi _{iS}(0)\rangle \}$\ is an initial basis of the system such that the 
$|\psi _{iS}(0)\rangle $ are macroscopically distinguishable to each other
(for a condition of macroscopic distinguishability, see \cite{Omnes-1994}, 
\cite{Omnes-1999}). In many models it can be shown that 
\begin{equation}
|\Psi (t)\rangle =\sum_{i}a_{i}|\psi _{iS}(t)\rangle \,|\mathcal{E}%
_{i}(t)\rangle  \label{3-32}
\end{equation}
where the $|\mathcal{E}_{i}(t)\rangle $ are the non-orthonormal states of
the environment. If the degrees of freedom of the environment are traced
over, we obtain the reduced state of the system $S$: 
\begin{equation}
\rho _{S}(t)=Tr_{E}|\Psi (t)\rangle \langle \Psi (t)|=\sum_{ij}a_{i}%
\overline{a_{j}}\,|\psi _{iS}(t)\rangle \langle \psi _{jS}(t)|\,\langle 
\mathcal{E}_{j}(t)\rangle |\mathcal{E}_{i}(t)\rangle  \label{3-33}
\end{equation}
In different examples it can be proved that, when the environment has many
degrees of freedom, for $t\rightarrow \infty $, the states of the
environment approach orthogonality, $\langle \mathcal{E}_{j}(t)|\mathcal{E}%
_{i}(t)\rangle \rightarrow \delta _{ji}$, and $\rho _{S}(t)$ becomes
diagonal in the so-called \textit{moving pointer} basis $\left\{ |\psi
_{iS}(t)\rangle \right\} $. The characteristic time of this process is the 
\textit{decoherence time} $t_{DS}$ \textit{of }$S$, which turns our to be
extremely short. Therefore, for $t\gg t_{DS}$, $\rho _{S}(t)$ results 
\begin{equation}
\rho _{S}(t)=\sum_{i}|a_{i}|^{2}|\psi _{iS}(t)\rangle \langle \psi _{iS}(t)|
\label{3-34}
\end{equation}
and follows its non-unitary evolution as a diagonal state, up to reach the
final stable state $\rho _{S*}$.

\subsection{Closed systems}

SID can be conceived as a closed-systems approach because it selects the
relevant observables without partitioning the closed system $U$, but in
terms of a different criterion.

Let us consider a quantum system endowed with a Hamiltonian $H$ with
continuous spectrum: $H|\omega \rangle =\omega |\omega \rangle $, $\omega
\in [0,\infty )$.

\textbf{Step 1: }A generic observable of the system reads 
\begin{equation}
O=\int_{0}^{\infty }\int_{0}^{\infty }\widetilde{O}(\omega ,\omega ^{\prime
})|\omega \rangle \langle \omega ^{\prime }|\,d\omega d\omega ^{\prime }
\label{3-35}
\end{equation}
where $\widetilde{O}(\omega ,\omega ^{\prime })$ is any distribution. The
relevant observables $O^{R}$ are those whose components are given by 
\begin{equation}
\widetilde{O^{R}}(\omega ,\omega ^{\prime })=O(\omega )\,\delta (\omega
-\omega ^{\prime })+O(\omega ,\omega ^{\prime })  \label{3.36}
\end{equation}
where $O(\omega ,\omega ^{\prime })$ is a regular function. Then, these
relevant observables $O^{R}$ belong to $\mathcal{O}_{VH}=\mathcal{O}^{R}$,
which we have called van Hove space (see \cite{van Hove-1}, \cite{van Hove-2}%
), and they read 
\begin{equation}
O^{R}=\int_{0}^{\infty }O(\omega )|\omega )\,d\omega +\int_{0}^{\infty
}\int_{0}^{\infty }O(\omega ,\omega ^{\prime })|\omega ,\omega ^{\prime
})\,d\omega d\omega ^{\prime }  \label{3-4}
\end{equation}
where $|\omega )=|\omega \rangle \langle \omega |$, $|\omega ,\omega
^{\prime })=\,|\omega \rangle \langle \omega ^{\prime }|$, and $\,\left\{
|\omega ),|\omega ,\omega ^{\prime })\right\} $ is a basis of $\mathcal{O}%
_{VH}$. In turn, states are represented by linear functionals belonging to $%
\mathcal{O}_{VH}^{\prime }$, the dual of $\mathcal{O}_{VH}$, and they read 
\begin{equation}
\rho =\int_{0}^{\infty }\rho (\omega )(\omega |\,d\omega +\int_{0}^{\infty
}\int_{0}^{\infty }\rho (\omega ,\omega ^{\prime })(\omega ,\omega ^{\prime
}|\,d\omega d\omega ^{\prime }  \label{3-41}
\end{equation}
where $\left\{ (\omega |,(\omega ,\omega ^{\prime }|\,\right\} $ is the
basis of $\mathcal{O}_{VH}^{\prime }$, that is, the cobasis of $\left\{
|\omega ),|\omega ,\omega ^{\prime })\right\} $. Under the usual
requirements ($\rho (\omega )$ real, positive and normalized), $\rho $
belongs to a convex space $\mathcal{S}\subset \mathcal{O}_{VH}^{\prime }$.

\textbf{Step 2: }The expectation value of any observable $O^{R}\in \mathcal{O%
}_{VH} $ in the state $\rho (t)\in \mathcal{S}$ can be computed as (see \cite%
{CL-PRA-2000}, \cite{CL-IJTP-2000}, \cite{CO-IJTP}, \cite{Cast-2004}, \cite%
{COrd}): 
\begin{equation}
\langle O^{R}\rangle _{\rho (t)}=\int_{0}^{\infty }\overline{\rho (\omega )}%
O(\omega )\,d\omega +\int_{0}^{\infty }\int_{0}^{\infty }\overline{\rho
(\omega ,\omega ^{\prime })}O(\omega ,\omega ^{\prime })\,e^{i\frac{\omega
-\omega ^{\prime }}{\hbar }t}\,d\omega d\omega ^{\prime }  \label{3-5}
\end{equation}

\textbf{Step 3: }When the function $\overline{\rho (\omega ,\omega ^{\prime
})}O(\omega ,\omega ^{\prime })$ is regular (precisely, when it is $\mathbb{L%
}_{1} $ in variable $\nu =\omega -\omega ^{\prime }$), the Riemann-Lebesgue
theorem can be applied to eq.(\ref{3-5}). As a consequence, the second term
vanishes and $\langle O^{R}\rangle _{\rho (t)}$ converges to a stable value: 
\begin{equation}
\langle O^{R}\rangle _{\rho (t)}\longrightarrow \langle O^{R}\rangle _{\rho
_{*}}=\int_{0}^{\infty }\overline{\rho (\omega )}O(\omega )\,d\omega
\label{3-6}
\end{equation}
where $\rho _{*}$ is diagonal in the eigenbasis of the Hamiltonian. The
characteristic time of this process is the \textit{relaxation time }$t_{RU}$ 
\textit{of the whole system }$U$, that is, the time that the system needs to
reach a state very close to the decohered equilibrium state.

Since in this case the whole system $U$ is not partitioned into $S$ and $E$,
the concept of moving pointer basis, defined in relation to a basis $\{|%
\mathcal{E}_{i}(t)\rangle \}$ of the environment, find no conceptual meaning
in the SID approach.

\section{Conditions for decoherence}

\subsection{Open systems}

In the context of the EID approach, the fast convergence of $\rho _{S}(t)$
has been obtained in several models. The paradigmatic example is the case of
a two-states system $S$ strongly coupled with an environment $E$ composed of
a large number $N$ of non-interacting particles. In this situation, the
environment behaves as a system with states $|\mathcal{E}_{i}(t)\rangle $,
and it can be proved that, when $N\rightarrow \infty $, $\langle \mathcal{E}%
_{i}(t)|\mathcal{E}_{j}(t)\rangle \rightarrow 0$ for $t\rightarrow \infty $:
the environmental states approach orthogonality and, as a consequence, $\rho
_{S}(t)$ approaches diagonality in the moving pointer basis in an extremely
short decoherence time $t_{DS}$.

>From the analysis of the models studied with this theoretical framework, it
can be concluded that environment-induced decoherence requires: (i) a
significant interaction between $S$ and $E$, and (ii) an environment $E$
with a huge number of degrees of freedom. Many physically relevant models
fulfill these conditions: in these cases the emergence of classical behavior
can be explained by the EID approach.

\subsection{Closed systems}

On the basis of the theoretical account of the SID approach, it is clear
that self-induced decoherence strictly obtains when the Hamiltonian has a
continuous spectrum. Nevertheless, the process also leads to decoherence in
quasi-continuous models, that is, discrete models where (i) the energy
spectrum is quasi-continuous, i.e., has a small discrete energy spacing, and
(ii) the functions of energy used in the formalism are such that the sums in
which they are involved can be approximated by Riemann integrals. This
condition is rather weak: the overwhelming majority of the physical models
studied in the literature on dynamics, thermodynamics, quantum mechanics and
quantum field theory are quasi-continuous, and the well-known strategy for
transforming sums in integrals is applied.

It is interesting to note that the selection of the relevant observables in
SID is also a very weak restriction. In fact, the observables not belonging
to the space $\mathcal{O}_{VH}$ are not experimentally accessible and, for
this reason, in practice they are always approximated, with the desired
precision, by regular observables for which the approach works
satisfactorily (for a full argument, see \cite{CO-Studies}).

\section{Relaxation and decoherence times}

Up to this point we have considered the convergence of the expectation
values to their final values. Now we will consider the characteristic times
involved in the processes.

\subsection{Open systems}

In several models studied by the EID approach, the decoherence time $t_{DS}$
of an open subsystem $S$ in interaction with its environment $E$ turns out
to be the relaxation time $t_{RS}$ of the system $S$ multiplied by a
macroscopicity coefficient $M$: 
\begin{equation}
t_{DS}=M\;t_{RS}  \label{4-1}
\end{equation}
For instance, in eq.(47) of \cite{Paz-Zurek} or in eq.(3.136) of \cite{Joos}%
, $M=\left( \frac{\lambda _{DB}}{L_{0}}\right) ^{2}$, where $\lambda _{DB}$
is the de Broglie length and $L_{0}$ is a macroscopic characteristic length.
In turn, in page 51 of \cite{Paz-Zurek}, $M=\left( \frac{\Delta x}{2L_{0}}%
\right) ^{2}$, where $\frac{\Delta x}{2L_{0}}$ is the ratio between a
microscopic and a macroscopic characteristic lengths.

Of course, the interaction between $S$ and $E$ is necessary to obtain a
finite relaxation time $t_{RS}$ since, with no interaction, $S$ and $E$ are
free evolving systems and $t_{RS}$ is infinite: in this case $t_{DS}$ is
also infinite and decoherence is only nominal. In the physically relevant
models studied by the EID approach, $t_{RS}$ is finite; thus, $t_{DS}$ is
extremely short since the macroscopicity ratios $\frac{\lambda _{DB}}{L_{0}}$
or $\frac{\Delta x}{2L_{0}}$ are extremely small (e.g. $10^{-20}$, see \cite%
{Paz-Zurek}). Therefore, $t_{DS}\ll $ $t_{RS}$.

\subsection{Closed systems}

According to the SID approach, decoherence is an irreversible evolution that
decays as $e^{-\frac{\gamma }{\hbar }t}$, where $\gamma $ is the imaginary
part of the pole closer to the real axis of the Hamiltonian resolvent: 
\begin{equation}
t_{RU}=\frac{\hbar }{\gamma }  \label{4-3}
\end{equation}
Of course, if the Hamiltonian has no poles, the closed system $U$ behaves as
a free evolving system. In this case, the system does not approach to a
final equilibrium state: its relaxation time $t_{RU}$ is infinite and, a
fortiori, decoherence is only nominal.

When the interactions introduce poles in the Hamiltonian, it can be
expressed as $H=H_{0}+V$, where $H_{0}$ is the free Hamiltonian and $V$ is
the interaction Hamiltonian containing the poles. It can be proved (see \cite%
{CO-PR(DT)}) that, in physically relevant cases, $\gamma \sim V$ and,
therefore, $t_{RU}\sim \hbar /V$.

For microscopic systems, $V$ can be estimated of the order of 1 e-V (a
natural energy scale for quantum atomic interactions, see e.g. \cite{Kuyatt}%
); then, the relaxation time $t_{RU}$ is very short, $\sim 10^{-15}s$ (see 
\cite{CO-PR(DT)}). For macroscopic systems, $V$ can be computed as $V=NV_{i}$%
, where $N$ is the number of subsystems (particles) of $U$, and each $V_{i}$
is the interaction between each subsystem and the rest of subsystems; in
this case, $t_{RU}\sim \hbar /NV_{i}$. If we consider again that all the $%
V_{i}$ are of the order of 1 e-V, and that $N=10^{24}$ (a macroscopic body
of 1 mol), the relaxation time $t_{RU}$ is fantastically short, $\sim
10^{-39}s$.

\subsection{Comparing both results}

In order to compare the results obtained in both cases, we have to partition
the whole system $U$ studied by SID into an open system $S$ and an
environment $E$. Now the Hamiltonian has to be expressed as: 
\begin{equation}
H=H_{0}+V=H_{0}+V_{SE}+V_{E}  \label{4-4}
\end{equation}
where $V_{SE}$ represents the interaction between $S$ and $E$, and $V_{E}$
represents the interactions of the subsystems (particles) of $E$ among
themselves.

If both $V_{SE}$ and $V_{E}$ contribute with poles, the relaxation time $%
t_{RU}$ of the whole system $U$ will be computed with the $\gamma $
corresponding to the pole closer to the real axis. Of course, if one of the
interactions is zero, the corresponding $V$ has no complex poles (can be
conceived as having a pole on the real axis), and the corresponding $\gamma $
is zero. Therefore, in this case the time $t_{RU}$, which has to be computed
with the pole closer to the real axis, is infinite: as expected, with no
interactions in a subsystem of the whole system $U$, the relaxation of $U$
is only nominal and, as a consequence, its decoherence is also only nominal.

An interesting situation is the case where both interactions have very
different strengths, in particular, $V_{SE}\gg V_{E}$. In this case, a
two-times evolution can be described (see \cite{CO-PR(DT)} for details):

\begin{enumerate}
\item Since $V_{SE}\gg V_{E}$, in a first step we can neglect $V_{E}$ and
consider the Hamiltonian $H^{(1)}=H_{0}+V_{SE}$. With this Hamiltonian we
can compute a relaxation time $t_{R}^{(1)}=\hbar /\gamma _{SE}$, where $%
\gamma _{SE}$ is the imaginary part of the pole (or of the pole closer to
the real axis) of $V_{SE}$. If the formalism of SID is applied to this case,
for times $t\gg t_{R}^{(1)}$, the state $\rho _{*}^{(1)}$ so obtained can be
considered diagonal for all practical purposes.

\item But since the whole system has not reached its final equilibrium state
yet, after the first period where $V_{SE}$ is dominant, for times $t\gg
t_{R}^{(1)}$, $V_{E}$ becomes relevant. In this situation, the total
Hamiltonian will be $H^{(2)}=H^{(1)}+V_{E}$, and the relaxation time $%
t_{R}^{(2)}=\hbar /\gamma _{E}$ can be computed, where $\gamma _{E}$ is the
imaginary part of the pole (or of the pole closer to the real axis) of $%
V_{E} $. Again, if the formalism of SID is applied, for times $t\gg
t_{R}^{(2)}$, we will obtain the state $\rho _{*}=\rho _{*}^{(2)}$, now
completely diagonal.
\end{enumerate}

>From this two-times evolution. we can see that:

\begin{itemize}
\item Since $\gamma _{SE}\sim V_{SE}$ and $\gamma _{E}\sim V_{E}$, when $%
V_{SE}\gg V_{E}$, $\gamma _{E}$ is the pole closer of the real axis by means
of which the relaxation time $t_{RU}$ of the whole system $U$ has to be
computed. Therefore, $t_{RU}=t_{R}^{(2)}$.

\item Since $t_{R}^{(1)}$ is computed only in terms of the interaction
between $S$ and $E$, it can be conceived as the relaxation time $t_{RS}$ of
the open system $S$: $t_{RS}=t_{R}^{(1)}$.

\item Since $V_{SE}\gg V_{E}$, then $t_{RS}=t_{R}^{(1)}\ll t_{R}^{(2)}=t_{RU}
$: as expected, the relaxation time $t_{RU}$ of the whole system $U=S\cup E$
will be much longer than the relaxation time $t_{RS}$ of the open system $S$%
: $t_{RU}\gg t_{RS}$. In other words, the time that a whole system needs to
reach the decohered state of equilibrium is much longer than the time needed
by a small subsystem strongly coupled with the rest of the degrees of
freedom. In turn, the relaxation time $t_{RS}$ that the system $S$ needs to
reach the decohered equilibrium is much longer than its decoherence time $%
t_{DS}$, that is, the time at which the state of $S$ becomes diagonal in the
moving pointer basis. $t_{RS}\gg t_{DS}$. As a consequence, $t_{RU}\gg
t_{RS}\gg t_{DS}$.

\item It is worth noting that the system $S$ may decohere and relax even in
the case that the subsystems of $E$ do not interact to each other. In this
case, the interaction $V_{E}$ is zero and, as explained above, $t_{R}^{(2)}$ 
$=t_{RU}$ is infinite: the whole system $U$ does not relax to an equilibrium
state and, as a consequence, it does not decohere. Nevertheless, the
relaxation time $t_{R}^{(1)}=t_{RS}$ can still be computed and, for a strong
interaction between $S$ and $E$, it will be extremely short. In turn, the
system $S$ may decohere in a decoherence time $t_{DS}\ll t_{RS}$. This means
that, even when the whole composite system does not decohere, one of its
subsystems strongly coupled with the remaining degrees of freedom may
decohere extremely fast.
\end{itemize}

\section{The spin-bath model}

The spin-bath model is a very simple model that has been exactly solved in
previous papers (see \cite{Zurek-1982}). Let us consider that the system $%
S_{0}$ is a spin-1/2 particle with states $|0\rangle $ and $|1\rangle $. The
environment $E$ is composed of $N$ spin-1/2 particles $S_{i}$ with states $%
|\uparrow _{i}\rangle $ and $|\downarrow _{i}\rangle $. The
self-Hamiltonians of $S_{0}$ and $E$ are taken to be zero, and $S_{0}$
interacts with $E$ via the interaction Hamiltonian $H_{SE}$: 
\begin{equation}
H_{SE}=\frac{1}{2}(|0\rangle \langle 0|-|1\rangle \langle
1|)\sum_{i=1}^{N}g_{i}(|\uparrow _{i}\rangle \langle \uparrow
_{i}|-|\downarrow _{i}\rangle \langle \downarrow _{i}|)\bigotimes_{j\neq
i}^{N}I_{j}  \label{6-1}
\end{equation}
where $I_{j}$ is the identity operator corresponding to the particle $S_{j}$%
. Then, the total Hamiltonian is simply $H=H_{SE}$.

A pure state of $U=S_{0}\cup E$ can be written as 
\begin{equation}
|\psi _{0}\rangle =(a|0\rangle +b|1\rangle )\bigotimes_{i=1}^{N}(\alpha
_{i}|\uparrow _{i}\rangle +\beta _{i}|\downarrow _{i}\rangle )  \label{6-2}
\end{equation}
where $\alpha _{i}$ and $\beta _{i}$ are aleatory coefficients such that $%
|\alpha _{i}|^{2}+|\beta _{i}|^{2}=1$. Under the action of $H=H_{SE}$, the
state $|\psi _{0}\rangle $ evolves into 
\begin{equation}
|\psi (t)\rangle =a|0\rangle |\mathcal{E}_{0}(t)\rangle +b|1\rangle |%
\mathcal{E}_{1}(t)\rangle  \label{6-3}
\end{equation}
where 
\begin{equation}
|\mathcal{E}_{0}(t)\rangle =|\mathcal{E}_{1}(-t)\rangle
=\bigotimes_{i=1}^{N}(\alpha _{i}e^{ig_{i}t/2}|\uparrow _{i}\rangle +\beta
_{i}e^{-ig_{i}t/2}|\downarrow _{i}\rangle )  \label{6-4}
\end{equation}
and the corresponding density matrix will be $\rho (t)=|\psi (t)\rangle
\langle \psi (t)|$.

An observable $O\in \mathcal{O}$ of the composite system $U=S_{0}\cup E$ can
be expressed as 
\begin{equation}
O=(s_{00}|0\rangle \langle 0|+s_{01}|0\rangle \langle 1|+s_{10}|1\rangle
\langle 0|+s_{11}|1\rangle \langle 1|)\bigotimes_{i=1}^{N}(\epsilon
_{\uparrow \uparrow }^{(i)}|\uparrow _{i}\rangle \langle \uparrow
_{i}|+\epsilon _{\downarrow \downarrow }^{(i)}|\downarrow _{i}\rangle
\langle \downarrow _{i}|+\epsilon _{\downarrow \uparrow }^{(i)}|\downarrow
_{i}\rangle \langle \uparrow _{i}|+\epsilon _{\uparrow \downarrow
}^{(i)}|\uparrow _{i}\rangle \langle \downarrow _{i}|)  \label{6-5}
\end{equation}
where $s_{00}$, $s_{11}$, $\epsilon _{\uparrow \uparrow }^{(i)}$, $\epsilon
_{\downarrow \downarrow }^{(i)}$ are real numbers and $s_{01}=\overline{%
s_{10}}$, $\epsilon _{\uparrow \downarrow }^{(i)}=\overline{\epsilon
_{\downarrow \uparrow }^{(i)}}$ are complex numbers. Then, the expectation
value of $O$ in the state $|\psi (t)\rangle $ reads 
\begin{equation}
\langle O\rangle _{\psi (t)}=(|a|^{2}s_{00}+|b|^{2}s_{11})\Gamma _{0}(t)+2%
\func{Re} [a\overline{b}\,s_{10}\Gamma _{1}(t)]  \label{6-6}
\end{equation}
where 
\begin{eqnarray}
\Gamma _{0}(t) &=&\prod_{i=1}^{N}[|\alpha _{i}|^{2}\epsilon _{\uparrow
\uparrow }^{(i)}+|\beta _{i}|^{2}\epsilon _{\downarrow \downarrow }^{(i)}+%
\overline{\alpha _{i}}\,\beta _{i}\epsilon _{\uparrow \downarrow
}^{(i)}e^{-ig_{i}t}+\overline{(\overline{\alpha _{i}}\,\beta _{i}\epsilon
_{\uparrow \downarrow }^{(i)})}\,e^{ig_{i}t}]  \label{6-7} \\
\Gamma _{1}(t) &=&\prod_{i=1}^{N}[|\alpha _{i}|^{2}\epsilon _{\uparrow
\uparrow }^{(i)}e^{ig_{i}t}+|\beta _{i}|^{2}\epsilon _{\downarrow \downarrow
}^{(i)}e^{-ig_{i}t}+\overline{\alpha _{i}}\,\beta _{i}\epsilon _{\uparrow
\downarrow }^{(i)}+\overline{(\overline{\alpha _{i}}\,\beta _{i}\epsilon
_{\uparrow \downarrow }^{(i)})}]  \label{6-8}
\end{eqnarray}

\subsection{The open-system viewpoint}

The open-system viewpoint consists in considering one of the spin-1/2
particles as the proper system and the remaining particles as the
environment. Let us consider two cases.\medskip \bigskip

\textbf{Case a): }In the typical situation studied by the EID approach, the
proper system is $S_{0}$. Therefore, the relevant observables $%
O^{R}=O^{S_{0}}$ are obtained by making $\epsilon _{\uparrow \uparrow
}^{(i)}=\epsilon _{\downarrow \downarrow }^{(i)}=1,$ $\epsilon _{\uparrow
\downarrow }^{(i)}=0$: 
\begin{equation}
O^{S_{0}}=(s_{00}|0\rangle \langle 0|+s_{01}|0\rangle \langle
1|+s_{10}|1\rangle \langle 0|+s_{11}|1\rangle \langle
1|)\bigotimes_{i=1}^{N}I_{i}  \label{6-9}
\end{equation}
The expectation value of these observables is given by 
\begin{equation}
\langle O^{S_{0}}\rangle _{\psi (t)}=|a|^{2}s_{00}+|b|^{2}s_{11}+\func{Re} [a%
\overline{b}\,s_{10}r(t)]  \label{6-10}
\end{equation}
where $r(t)=\langle \mathcal{E}_{1}(t)|\mathcal{E}_{0}(t)\rangle $ and 
\begin{equation}
|r(t)|^{2}=\prod_{i=1}^{N}(|\alpha _{i}|^{4}+|\beta _{i}|^{4}+2|\alpha
_{i}|^{2}|\beta _{i}|^{2}\cos 2g_{i}t)  \label{6-11}
\end{equation}
Numerical results show that (see \cite{Max}), as $N$ increases, $|r(t)|$
quickly decays by several orders of magnitude. This means that the
interference is suppressed from the viewpoint of the observables that
``observe'' only the spin system $S_{0}$.\bigskip

\textbf{Case b):} Nevertheless, we can also decide to select the observables 
$O^{S_{j}}$ that ``observe'' just \textit{one} spin system $S_{j}$ of $E$ as
the relevant observables: 
\begin{equation}
O^{S_{j}}=I_{S_{0}}\otimes O_{j}\bigotimes_{i\neq j}I_{S_{i}}  \label{6-12}
\end{equation}
where $\epsilon _{\uparrow \uparrow }^{(j)}$, $\epsilon _{\downarrow
\downarrow }^{(j)}$, $\epsilon _{\uparrow \downarrow }^{(j)}$ are now
generic. The expectation value of these observables is given by 
\begin{equation*}
\langle O^{S_{j}}\rangle _{\psi (t)}=\langle \psi (t)|O_{S_{j}}|\psi
(t)\rangle =|a|^{2}(|\alpha _{j}|^{2}\epsilon _{\uparrow \uparrow
}^{(j)}+|\beta _{j}|^{2}\epsilon _{\downarrow \downarrow }^{(j)}+\alpha _{j}%
\overline{\beta _{j}}\,\epsilon _{\uparrow \downarrow }^{(j)}e^{-ig_{j}t}+%
\overline{\alpha _{j}}\beta _{j}\epsilon _{\downarrow \uparrow
}^{(j)}e^{ig_{j}t})+ 
\end{equation*}
\begin{equation}
|b|^{2}(|\alpha _{j}|^{2}\epsilon _{\uparrow \uparrow }^{(j)}+|\beta
_{j}|^{2}\epsilon _{\downarrow \downarrow }^{(j)}+\alpha _{j}\overline{\beta
_{j}}\,\epsilon _{\uparrow \downarrow }^{(j)}e^{ig_{j}t}+\overline{\alpha
_{j}}\beta _{j}\epsilon _{\downarrow \uparrow }^{(j)}e^{-ig_{j}t})
\label{6-13}
\end{equation}
>From this equation it is easy to see that $\langle O^{S_{j}}\rangle _{\psi
(t)}$ oscillates and, thus, it has no limit. Therefore, from the viewpoint
of the observables that ``observe'' only the spin system $S_{j}$ certainly
there is no decoherence. This is not a surprising result when we recall
that, in the Hamiltonian of eq. (\ref{6-1}), the spin systems of the
subsystem $E=\bigcup_{i}S_{i}$ are uncoupled to each other: each $S_{i}$
evolves as a free system and, for this reason, $E$ is unable to reach a
final stable state.

\subsection{The closed-system viewpoint}

>From the closed-system viewpoint of the SID approach, the system $%
U=S_{0}\cup E$ is considered as a whole. Then, the relevant observables are
given by eq. (\ref{6-5}) and their expectation values are computed as in eq.
(\ref{6-6}).

Numerical results (see \cite{Max}) show that, in the general case, the time
evolution of eq. (\ref{6-6}) does not lead to the suppression of the terms
in $\langle O\rangle _{\psi (t)}$ that are not diagonal in the energy
eigenbasis; as a consequence, the system $U$ does not relax nor decohere
from the perspective of SID. Schlosshauer (\cite{Max}) interprets this
result as a shortcoming of the SID approach: SID would fail to explain the
phenomenon of decoherence in this model, correctly described by the EID
approach. According to the author, such a failure is due to the \textit{%
discrete} nature of the model, even under the condition of quasi-continuous
energy spectrum. On this basis, he concludes that SID is likely to fail in
other systems composed of discrete subsystems.

When decoherence is understood in the context of the general framework
introduced here, it is easy to see that, in spite of the interesting
numerical simulations, Schlosshauer's interpretation of the results is
misguided, since it relies in comparing processes resulting from different
subsets of relevant observables. In fact, in a given system decoherence is
not a yes-no phenomenon, but a process relative to the relevant observables
chosen for the description. When this point is taken into account, the fact
that a system may decohere for certain subset of relevant observables and
may not decohere for a different subset turns out to be natural: in
particular, this is what happens in Cases a) and b) of the previous
subsection, both solved from the open-system perspective of EID.

Furthermore, when the relaxation time $t_{RU}$ computed as explained in
Section V is considered, the fact that the whole system $U$ will not
decohere from the SID approach, far from being a shortcoming of the
approach, is its necessary consequence. Let us recall that $t_{RU}$ is
obtained on the basis of the poles of the total Hamiltonian of the closed
system $U$. The spin-bath model is a paradigmatic example of the situation
considered in Subsection V.C and developed in paper \cite{CO-PR(DT)}, where $%
V_{SE}\gg V_{E}$. In particular, the model is a case where $V_{E}=0$, as
explained in that subsection: since the particles of the environment $E$ do
not interact to each other, $t_{RU}$ is infinite. Therefore, it can be
easily inferred, with no need of numerical simulations, that in this model
the whole system $U$ does not decohere. Nonetheless, this is not an obstacle
for the decoherence of the subsystem $S_{0}$, strongly coupled with the
remaining degrees of freedom: the corresponding relaxation time $t_{RS_{0}}$
can be computed from the interaction Hamiltonian $H_{SE}$, and the
decoherence time $t_{DS_{0}}$ will be even shorter: $t_{DS_{0}}\ll
t_{RS_{0}} $.

\section{Concluding Remarks}

In this paper we have proposed a general theoretical framework for
decoherence, which encompasses formalisms originally devised to deal just
with open or with closed systems. When decoherence is understood in this
framework, the conceptual difficulties of the EID program turn out to be not
as serious as originally supposed. In fact:

\begin{itemize}
\item Closed quantum systems may decohere; furthermore, in spite of the fact
that EID focuses on open systems, it can also be formulated from the
perspective of the whole composite system and, in this case, meaningful
relationships between the behavior of the whole system and the behavior of
its subsystems can be explained.

\item The ``defining systems'' problem is simply dissolved by the fact that
the splitting of the closed system into an open subsystem and an environment
is just a way of selecting the relevant observables of the closed system.
Since there are many different sets of relevant observables depending on the
observational viewpoint adopted, the same closed system can be decomposed in
many different ways: each decomposition represents a decision about which
degrees of freedom are relevant and which can be disregarded in any case.
Since there is no privileged or ``essential'' decomposition, there is no
need of an unequivocal criterion to decide where to place the cut between
``the'' system and ``the'' environment.
\end{itemize}

>From this theoretical perspective, decoherence is not an ``absolute''
phenomenon which occurs or does not occur in a given system. On the
contrary, decoherence is relative to the relevant observables selected in
each particular case: there is not a privileged set of relevant observables.
The only essential physical fact is that, among all the observational
viewpoints that may be adopted to study a quantum system, some of them
determine subsets of relevant observables for which the system decoheres.

As a consequence, the formalisms of decoherence for open and for closed
systems are not rival or alternative, but they cooperate in the
understanding of the same physical phenomenon. Therefore, the results
obtained in both cases turn out to be relevant: for instance, the large
amount of experimental confirmations of EID (see \cite{Joos}), the complete
description of the classical limit of quantum mechanics (\cite{Cast-2004}, 
\cite{CG-CL}, \cite{CO-IJTP}, \cite{CO-PS}) and the study of the role of
complexity in decoherence (\cite{Cast-2005}, \cite{Cast-2006}, \cite{CO-CSF}%
) in the case of SID, and the meaningful relations between the decoherence
times computed by EID and the relaxation times computed by SID (\cite%
{CO-PR(DT)}) can be all retained as important acquisitions in the new
general framework.

\section{Acknowledgements}

We are very grateful to Roland Omn\`{e}s and Maximilian Schlosshauer for
many comments and criticisms. This research was partially supported by
grants of the University of Buenos Aires, CONICET, and FONCYT of
Argentina.

\end{document}